# Sequence to Point Learning Based on Bidirectional Dilated Residual Network for Non-Intrusive Load Monitoring

Ziyue Jia[a,b], Linfeng Yang[a,b,*], Zhenrong Zhang[a,b], Hui Liu[c,d], Fannie Kong[c,d]

[a]School of Computer Electronics and Information, Guangxi University, Nanning 530004, China
[b]Guangxi Key Laboratory of Multimedia Communication and Network Technology, Guangxi University, Nanning 530004, China
[c]School of Electrical Engineering, Guangxi University, Nanning 530004, China
[d]Guangxi Key Laboratory of Power System Optimization and Energy Technology, Guangxi University, Nanning 530004, China

*Abstract*—Non-Intrusive Load Monitoring (NILM) or Energy Disaggregation (ED), seeks to save energy by decomposing corresponding appliances power reading from an aggregate power reading of the whole house. It is a single channel blind source separation problem (SCBSS) and difficult prediction problem because it is unidentifiable. Recent research shows that deep learning has become a growing popularity for NILM problem. The ability of neural networks to extract load features is closely related to its depth. However, deep neural network is difficult to train because of exploding gradient, vanishing gradient and network degradation. To solve these problems, we propose a sequence to point learning framework based on bidirectional (non-casual) dilated convolution for NILM. To be more convincing, we compare our method with the state of art method—Seq2point (Zhang) directly and compare with existing algorithms indirectly via two same datasets and metrics. Experiments based on REDD and UK-DALE data sets show that our proposed approach is far superior to existing approaches in all appliances.

## 1. Introduction

With the gradual deterioration of the global climate and the increasing scarcity of renewable energy, energy conservation and emission reduction have become an important issue that cannot be ignored. To encourage people to conserve electricity, a series of regulations and tiered pricing policies were published worldwide. However, it is not enough only to rely on the government's macro-control policy, how to cultivate consciousness of electricity-consumption and lead people develop the related habit still needs to be further discussed. In order to enable home users to understand the energy consumption of household appliances more clearly and accurately and optimize the power consumption management, a power consumption visualization technology, load monitoring, is increasingly becoming a critical topic in the academic and industry. Generally, load monitoring is divided into intrusive load monitoring and non-intrusive load monitoring. Intrusive load monitoring refers to a technology that records the use of electricity by installing sensors in each electrical appliance, which is characterized by high accuracy, high cost and low feasibility of popularization. Another method is non-intrusive load monitoring (NILM), firstly proposed as load disaggregation by Hart in 1992 [1], records aggregate power-reading by installing a sensor in the household center meter, estimating each appliance power usage by using a series of algorithms, so as to obtain overall law of the power consumption. NILM can not only help reduce the energy consumption of household users by 15%, bring economic benefits to household users and power grid companies, but also make a significant contribution to global energy conservation and emission reduction [2,3]. Compared with intrusive load monitoring, NILM is used in the mainstream with the advantages of small economic input and strong practicability [4]. Therefore, researchers pay more attention to NILM in recent years. In the past decade, mathematical optimization methods and pattern recognition were the major methods in studying load disaggregation. Hart et al. [1] used combinational optimization (CO) to divide each device into multiple states, and each state has a corresponding power consumption. Ahmadi et al. [5] proposed eigenloads approach where it efficiently distinguishes highly similar load features. Chang et al. [6] and Lin et al. [7] applied a particle swarm optimization (PSO) to simultaneously decompose the aggregate power consumption into each corresponding appliance, but there is a big error in metric. Pipa et al. [8] used a sparse optimization (SO) for ED to improve the accuracy of disaggregation. Lin et al. [9] proposed a novel load identification based quadratic programming (QP) [9], which has high identification accuracy and anti-interference capability. Kim et al. [10] proposed an ED method based multi-factor hidden Markov (MFHMM). Kolter et al. [11], Zhong et al. [12], Parson et al. [13] and Bonfigli [14] used additive factorial hidden Markov (AFHMM) to solve ED problem. With the advent of the big data and the rise of machine learning, pattern recognition (PR) was applied to NILM which includes k-nearest neighbor regression(KNN) [15], support vector machine(SVM) [16], Adaboost algorithm [17], clustering [18,19], Bayes [20], fuzzy system [21,22]. Experiments mentioned before having proven that PR improved disaggregation accuracy, but it requires manually extract features from a large amount of data such that difficult to catch in practice.

With the purpose of improving accuracy and practicality, researchers of non-intrusive load have shifted their focus to apply deep learning in NILM. Deep learning is an end-to-end technique, where neural network can extract features from data automatically rather than manual work. In 2015, Kelly et al. [23] proposed to combine Convolutional Neural Network (CNN), Recurrent Neural Networks (RNNs) and Denoising Autoencoder (DAE) with Sequence-to-Sequence Learning (Seq2seq). Seq2seq means that network receives a sequence of data and output a sequence of prediction data after being mapped. To overcome the vanishing gradient problem, Kelly et al. chose Long Short-Term Memory (LSTM) instead of traditional RNN. Under the UK-DALE [24] data set, the performance of the three networks proposed by Kelly et al., exceeds Hart's combinatorial optimization algorithm and the Factorial HMM. Similarly, Mauch et al. [25] combined sequence to sequence learning and LSTM as well. The only difference was the way of input-output where this network used the aggregate power consumption value at a specific time point to predict the consumption of target appliance at the same time point. In 2018, Zhang et al. [26] achieved the state-of-the-art results by combining CNN and Seq2point. Seq2point means the network receives a sequence of data and outputs a prediction point after being mapped. The Seq2point (Zhang) would input a window of aggregate data, predicting the midpoint of the same window of appliance consumption. Zhang et al. compared their Seq2seq and Seq2point methods with Kelly's work under the two-common metrics of average absolute error (MAE) and total signal error (SAE), and found that their performance were far superior to Seq2seq(Kelly) in both the UK-DALE dataset and the REDD dataset [27]. Under the UK-DALE dataset, compared with Seq2seq (Kelly), overall MAE of Seq2seq (Zhang) decreased by 81% and SAE by 89%, while Seq2point (Zhang) has a decrease of 84% on MAE and 92% on SAE. Overall Seq2point (Zhang) also performed better than Seq2seq (Zhang) on the REDD dataset, with MAE down 11% and SAE down 24%. In 2018, Based on the LSTM network for NILM proposed by Kelly, Krystalakos et al. [28] proposed an online energy disaggregation method on GRU neural network, which input a window of past aggregate data to infer the last point of the window. Compared with the Seq2seq (Kelly), it contains the following two advantages: 1) 60% reduction of trainable parameters with the same performance [29]. 2) online disaggregation is achievable when use the previous sample points to predict the current sample points.

In this paper, we propose bidirectional dilated residual network to sequence-to-point load disaggregation, outperforming existing approaches in all appliances. Therefore, it is very important to keep the sliding window size enough length to contain as much the appliance activations as possible, because the increasing of window size can decrease the unbalanced distribution of sequential data. Nevertheless, the increasing size of sliding window requires the convolution kernel has a larger receptive field and increases memory difficulty of long sequential data. Hence, bidirectional dilated convolution is introduced. Compared with standard convolution and dilated convolution, it enlarges the receptive field and decrease computation cost. By stacking deeper layers, it also makes the receptive field larger, while more load relations are constructed to more complex load features gradually. The more complex features extract, the better performance. However, deep layers lead to vanishing gradient, exploding gradient and networking degradation. In order to address these problems, we introduce residual block from temporal convolution network (TCN). Different with the original TCN residual block, we use bidirectional (non-casual) dilated convolution and batch normalization instead of dilated convolution and weight normalization. Within the residual block, batch normalization [30] and activation function ReLU are deployed to solve vanishing gradient and exploding gradient, while residual connection [31] is used to prevent networking degradation. Experimental results founded on two open benchmark data sets, REDD and UKDALE (2015), point our method outperforms state-of-the-art Seq2seq (Zhang) and Seq2point (Zhang) in all appliances.

## 2. Single Channel Blind Source Separation (SCBSS)

In this section, we will introduce SCBSS and the research areas it covers, so that methods proposed from one domain can be easily solved in another related domain. Imagine a "cocktail party" where the sounds of conversation and music mix. For guests at a party, they can easily understand what their partner is saying even in a noisy environment. The phenomenon of selecting one voice from mixed sounds while ignoring the others is known as the cocktail effect. How to separate the source signal from this observed mixed signal is called the SCBSS problem. Suppose a single sensor receives observation signal $Y(t)$ at time $t$ of discrete sampling, which is composed of source signals $X_i(t)$, SCBSS problem can be expressed as formula:

$$Y(t) = \sum_{i=1}^{I} X_i(t) + n(t) \quad (1)$$

where $X_i(t)$ is $i$th source signal, $I$ is total number of source signals and $n(t)$ is gaussian noise.

The definition of NILM problem is to predict respective appliances power consumption from an aggregate power reading of whole house measured by center meter. We apply the definition of NILM to the formulation (1), treat $i$th appliance power as $X_i(t)$ and aggregate power reading as $Y(t)$. It is easy to see that NILM problem belongs to SCBSS problem. What' more, denoising issues such as speech and audio can be viewed as SCBSS problem as well. To be more detailed, formula (2) can be converted and inferred from formula (1).

$$Y(t) = \sum_{i=1}^{I-1} X_i(t) + n(t) + X_I(t) \quad (2)$$

In denoise domain, $Y(t)$ is seen as a mixed signal. $X_I(t)$ is clean signal, $\sum_{i=1}^{I-1} X_i(t) + n(t)$ is considered as a whole parameter, that is noise.

In brief, SCBSS problem includes NILM problem, denoising issues, etc. Each subsequence can be solved by referring to the other solutions or strategies under SCBSS problem.

## 3. Proposed Neural Network Architectures

Our model is inspired by exploring denoising problem in Wavenet [32] since both NILM problem and denoising problem belong to SCBSS problem. Additionally, this paper is lighted by TCN [33] as well. Our model is further discussed in this section.

### 3.1 Bidirectional(non-casual) vs casual dilated network.

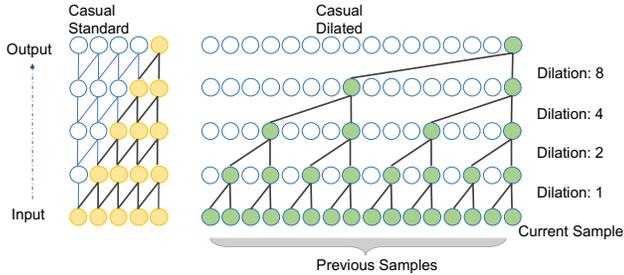

Fig. 1. Casual standard convolution(left) and Casual dilated(right) stacks.

This section will introduce you to casual standard convolution, casual dilated convolution, bidirectional dilated convolution, and then further explain why bidirectional dilated convolution is applied in our model. Unlike standard convolution, casual standard convolution uses the previous time step sample to predict the current result. Fig. 1 has two convolution stacks, where the causal standard convolution stack (left) and the causal dilated stack (right). They all have the same layers and filter length, but different dilation factors.

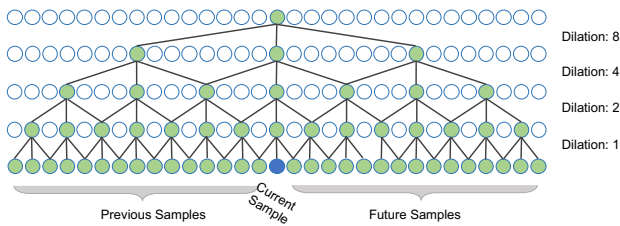

Fig. 2. Bidirectional (Non-casual) dilated convolution stacks.

For the causal standard convolution (left), the dilation factor remains 1 and the receptive field increases by 1 each layer. In order to obtain a larger receptive field, casual dilated is introduced. As shown in right of Fig. 1, the dilation factor for each layer is increased by a factor of 2, making the receptive field grow exponentially with depth. In NILM area, casual is not necessary and future samples are generally useful for improving predictions. As shown in Fig. 2, a transformation of dilated casual convolution pattern is called bidirectional dilated convolution. It eliminates causality to access the same number of samples in the past as samples in the future to make the prediction at center of receptive field, which result in larger receptive field and better prediction result. For more detail, as shown in Fig. 1, casual standard convolution and casual dilated stacks both have 4 layers and filter length 2 in each layer. The length of the receptive field of casual standard is 5, casual dilated is 16. However, as indicated in Fig. 2, bidirectional dilated convolution stacks have 4 layers and filter length 3 for each layer. The length of the receptive field is 31. Compared with casual dilated stacks (Fig. 2), the filter length of bidirectional dilated convolution increases 1 but the length of the receptive field is double. Bidirectional dilated convolution structure makes it very efficient to increase the length of receptive field.

### 3.2 Residual Block

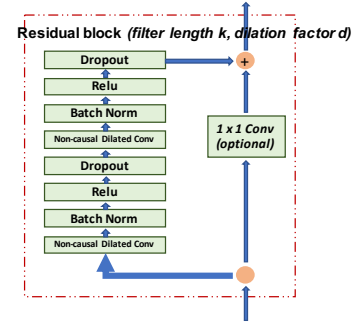

Fig. 3. Residual Block Structure

The depth of the neural network has an important influence on the performance of the model. With the increase of the depth of neural network, it can obtain a larger receptive field and extract more load features of different level, so that the neural network can theoretically obtain better performance. However, with the increase of the depth of neural network, the problems of vanishing gradient and explosion gradient arise make neural network difficult to train. In order to overcome the problems of vanishing gradient and explosion gradient, the activation function ReLU and batch normalization are introduced. In addition, degradation occurs when the neural network reaches a certain depth, which means that the accuracy will be saturated or even decreased if the depth of the neural network continues to increase. Experiments show that residual (skip) connection can solve the degradation problem. After adding the residual connection, the deeper network is not worse than the shallow network [31].

TCN residual block has 2 residual units, each of which contains dilated convolution, weight normalization [34], ReLU activation and dropout. Dropout is applied to prevent overfitting [35]. In addition, TCN residual block includes weight normalization, activation function ReLU and residual connection, which is a nice method to solve the vanishing gradient and degradation problem. The residual connection makes the deep network no worse than the shallow network, and its 1 x 1 convolution will be used when the input and output directions are different. Our TCN residual block is illustrated in Fig. 3. Unlike the original TCN residual block, we use bidirectional (non-casual) dilated convolution and batch normalization instead of dilated convolution and weight normalization to give the neural network a larger receptive field and better performance.

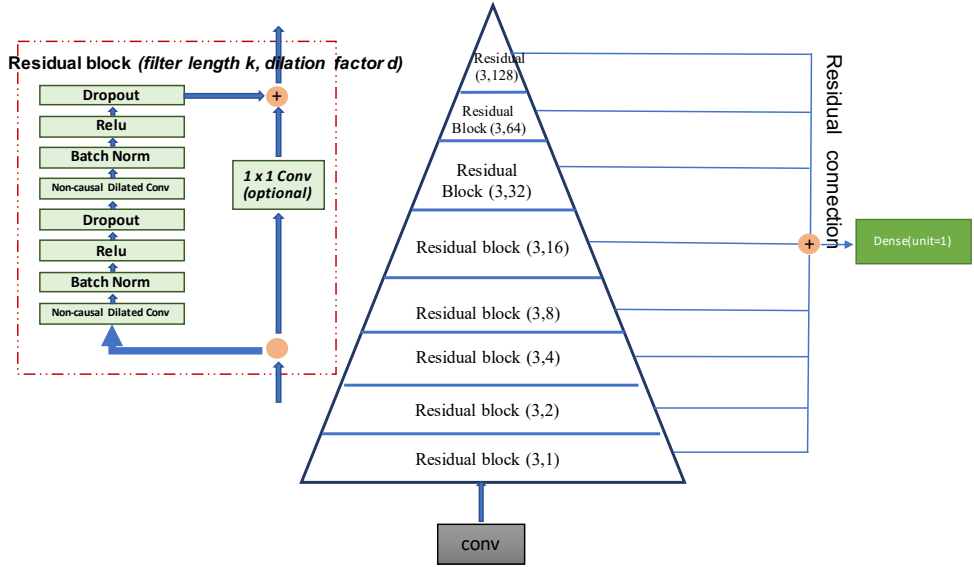

Fig. 4. Sequence to point learning based on bidirectional dilated residual network for NILM

## 3.3 Sequence to point learning based on bidirectional dilated residual network for NILM

Our method is illustrated in Fig. 4. It adopts a sliding window of the aggregate power as input to standard convolution for extracting low-level load features. The output of the first convolution is then progressively fed into the 8 residual blocks to extract the higher level of load features. The eight residual blocks (layers) form the triangle shown in the center of Fig. 4, derived from the bidirectional dilated convolution stacks shown in Fig. 2. The only difference is that each bidirectional dilated convolutional layer is replaced by the residual block. The residua block can be treated as enhancing bidirectional dilated convolution layer because it can solve the problem of vanishing gradient and explosion gradient. For each residual block, the number and length of filters are 128 and 3, respectively, but the dilation factor is increased by 2 times. Besides, each residual block has two consequences. One consequence is fed to next residual block for higher features, while another skips all subsequent layers by residual connection and then considered as an addition operation to extract the highest feature vector. Finally, dense layer maps the feature vector into a target point. In this paper, Adam optimizer algorithm and mean square error loss are applied. Our model contains 912,641 parameters in total.

## 4. Experiments

Zhang's experiments [26] show that their Seq2point method achieve the best performance compared with AFHMM [11], Seq2seq (Kelly) [23], and Seq2seq (Zhang) [26].

In this section, we compare our model with the Seq2point (Zhang) [26]. Our experiments are performed in Python 3 using TensorFlow and Keras. We train it on a machine with NVIDIA Tesla V100.

### 4.1 Data sets

Our experiments are founded on two real-world datasets, REDD and UK-DALE.

In REDD data set, total aggregate power consumption is recorded every 1 second and individual appliance power consumption is accounted for every 3 seconds. The data set contains measurements of six U.S. homes and 10 to 25 household appliances. It is very important that a neural network can generalize appliances to unseen houses since we do not have enough ground truth appliance data in the house when NILM is applied in real life. In experiment 1, we make use of the data of house 2-6 for training, and the data of household 1 for testing, just like the work of Seq2point (Zhang). Like Seq2point (Zhang), we only consider microwaves, fridge, dishwashers and washing machines as well.

On UK-DALE data set published in 2015, the aggregate power consumption and appliances power consumption were recorded in every 6 seconds. The data set contains 5 UK houses and measures of 4-54 appliances. Same as Seq2point (Zhang), we look at kettle, microwave, fridge, dish washer and washing machine. Experiment 2 is implemented on household 2 of UK-DALE data set. The first 80 % data of each family is training set and remaining 20% is testing set.

### 4.2 Data Preprocess

In this section, we describe how to prepare the data for the experiment. In our experiments, a sliding window of the aggregate power reading is used as the input sequence, and the midpoint of the window corresponding to the target device reading is used as the output. In order to eliminate the influence of different units or scales, it is transformed into dimensionless value and the data is normalized. We normalize all data by subtracting the mean and dividing by the standard deviation. These parameters and window lengths, as shown in Table 1, affect the experimental results, so we set them exactly same as Seq2point (Zhang) and Seq2seq (Zhang) for fair comparison. The window length is not optimal for our proposed method.

Table 1: Parameter Used for Each Appliance

| Appliance | Windows Length | Mean Power | Standard deviation of Power |
|---|---|---|---|
| Kettle | 599 | 700 | 1000 |
| Microwave | 599 | 500 | 800 |
| Fridge | 599 | 200 | 400 |
| Dish washer | 599 | 700 | 1000 |
| Washing machine | 599 | 400 | 700 |

*4.3 Performance evaluation*

Two metrics are selected as measurements to compare these methods: Mean Absolute Error (MAE) and Signal Aggregate error (SAE). MAE can be expressed as:

$$\text{MAE} = \frac{1}{T}\sum_{t=1}^{T}|\hat{x}_t - x_t|,$$

where $\hat{x}_t$ is the prediction of an appliance in time $t$ and $x_t$ is ground truth value. $T$ is the number of time points. MAE measure the mean of error in power at each time point. SAE can be defined as:

$$\text{SAE} = \frac{|\hat{r} - r|}{r},$$

where $\hat{r}$ and $r$ represent the predicted total energy consumption of an appliance and the ground truth one. SAE measures the total error in the energy within a period, which is accurate for reporting daily power consumption even if its consumption is inaccurate in every time point.

## 5. Result

Compared to Seq2seq (Zhang) and Seq2point (Zhang), as shown in Table 2, our method has fewer parameters than other methods. In addition, our approach is far superior to the others in the two experiments based on REDD or UK-DALE data sets.

Table 2: Total parameters for these methods

| Method | Total parameters |
|---|---|
| Seq2seq (Zhang) | 8,332,399 |
| Seq2point (Zhang) | 30,708,249 |
| Seq2point (This paper) | 912,641 |

The Experiment 1 is firstly performed on REDD data set and it is implemented on the same experiment environment with Zhang et al. paper [26], as described above. To be more convincing, the experiment result of Seq2seq (Zhang) and Seq2point (Zhang) are referred from their paper. Table 3 shows that our approach is superior to the others in terms of MAE and SAE metric. In comparison with Seq2seq (Zhang), our method improve performance and number was reduced by 28% overall and decreased microwave, fridge, dishwasher, and washing machine by 42%, 13%, 9%, and 44%, respectively on the MAE metric. On the SAE metric, the performance was improved, and overall number was reduced by 52%, while microwaves, fridge, dishwashers, and washing machines were decreased by 51%, 45%, 18%, and 91%, respectively. Compared with Seq2point (Zhang), our number was reduced overall by 19%, and by 31%, 5%, 12%, and 31%, respectively, for microwave, fridge, dishwasher, and washing machine on MAE metric. On the SAE metric, our method improved performance by decreasing to 37% overall, and fridge, dishwasher, and washing machine decreased by 65%, 20%, and 83%, respectively. The disaggregation of household 1 is described in Fig. 5, where the aggregate, ground truth and S2P represent aggregate consumption, real value of target appliance and prediction value of target appliance by S2P method, respectively. For the fridge disaggregate graph, At the first glance, it seems that our method does not predict well on total time steps since there is a big visual difference between ground truth value(red) and predict value(blue). However, in the Fig. 6, it is easy to be seen that our model can learn load signature well except for the peak ground truth value. Those peak ground truth values are greater than responding aggregate power values, but we could not find any paper point out the problem. After reviewing the REDD data set again, we identified the mistake from data set that those peak values are negligible because the number is very few. In general, our method also does a good job of predicting fridge, even if it does not look good visually. Fig. 6 shows that our model can efficiently extract load signatures and generalize appliances to house not seen during training.

To further validate our method, the experiment 2 based on UK-DALE data set are performed. Table 4 shows that our method also achieved the best results when comparing with Seq2point (Zhang). Compared with Seq2point (Zhang), our method improve performance and number was reduced by 33% overall and decreased kettle, microwave, fridge, dishwasher, and washing machine by 56%, 51%, 31%, 4% and 26%, respectively on the MAE metric. On the SAE metric, our method improved performance by decreasing to 38% overall, and kettle, microwave, fridge and dish washer decreased by 63%, 56%, 29% and 90%, respectively. Fig. 8 and Fig. 7 show that the overall and partial decomposition effects of our method and Seq2point (Zhang) respectively on household 2 of UK-DALE data set. In Fig. 8, by matching the predicted value of target appliance with the ground truth value of the target appliance, it is obvious that our prediction is closer to the real value than Seq2Point (Zhang). Fig. 7 also proves that our method fits the real value of all appliances well, which is significantly better than Seq2Point (Zhang).

Table 3: Performance comparison of algorithms on household 1 of REDD data set. Best results shown in bold.

| Error metric | Models | Microwave | Fridge | Dish washer | Washing machine | Overall |
|---|---|---|---|---|---|---|
| MAE | Seq2seq (Zhang) [26] | 33.272 | 30.63 | 19.449 | 22.857 | 26.552 |
| | Seq2point (Zhang) [26] | 28.199 | 28.104 | 20.048 | 18.423 | 23.693 |
| | Seq2point (this paper) | **19.455** | **26.801** | **17.665** | **12.763** | **19.171** |
| SAE | Seq2seq (Zhang) | 0.242 | 0.114 | 0.557 | 0.509 | 0.355 |
| | Seq2point (Zhang) | **0.059** | 0.18 | 0.567 | 0.277 | 0.27 |
| | Seq2point (this paper) | 0.118 | **0.063** | **0.454** | **0.048** | **0.171** |

Table 4: Performance comparison of algorithms on household 2 of UK-DALE data set. Best results shown in bold.

| Error metric | Models | Kettle | Microwave | Fridge | Dish washer | Washing machine | Overall |
|---|---|---|---|---|---|---|---|
| MAE | Seq2point (Zhang) | 4.881 | 2.686 | 11.807 | 3.633 | 5.509 | 5.703 |
| | Seq2point (this paper) | **2.16** | **1.305** | **8.136** | **3.49** | **4.063** | **3.831** |
| SAE | Seq2point (Zhang) | 0.121 | 0.251 | 0.056 | 0.021 | **0.096** | 0.109 |
| | Seq2point (this paper) | **0.045** | **0.111** | **0.040** | **0.002** | 0.143 | **0.068** |

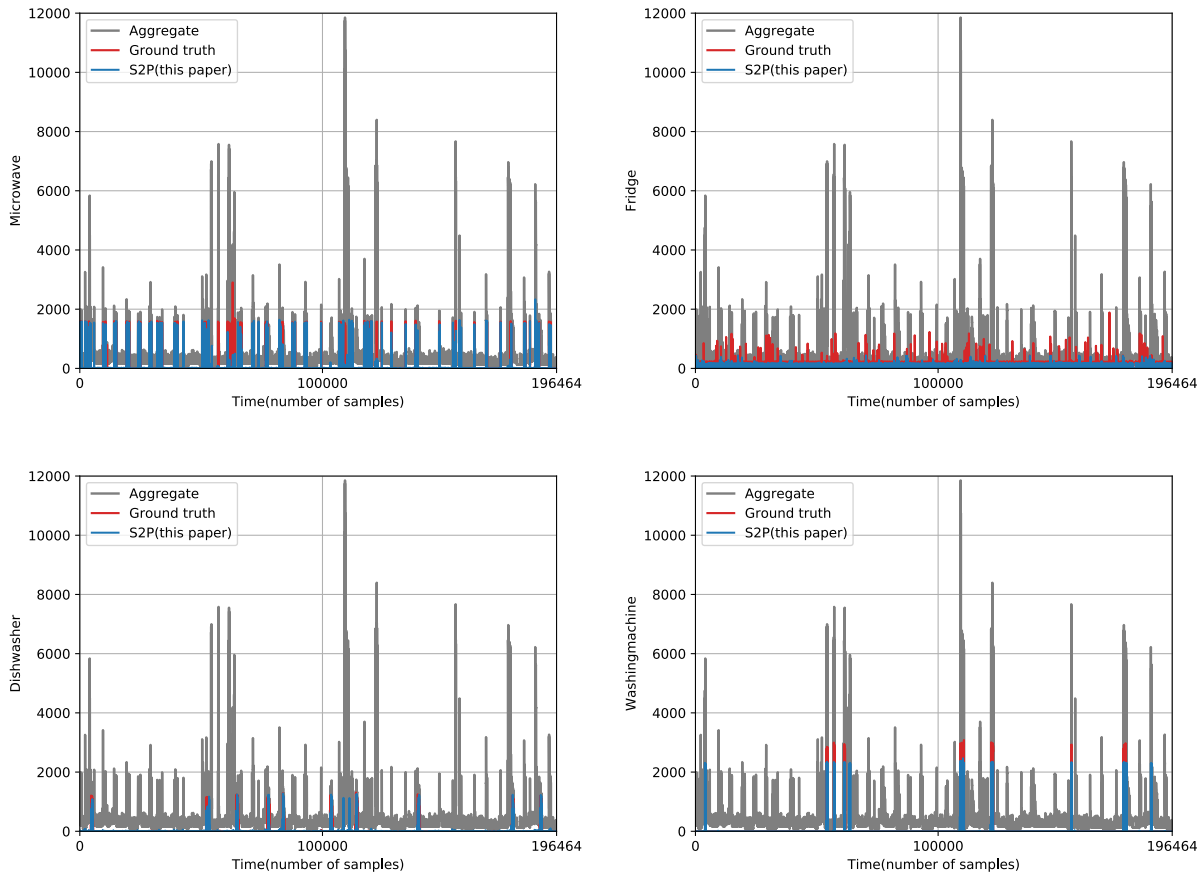

Fig. 5. Disaggregate result of our algorithm on REDD data set.

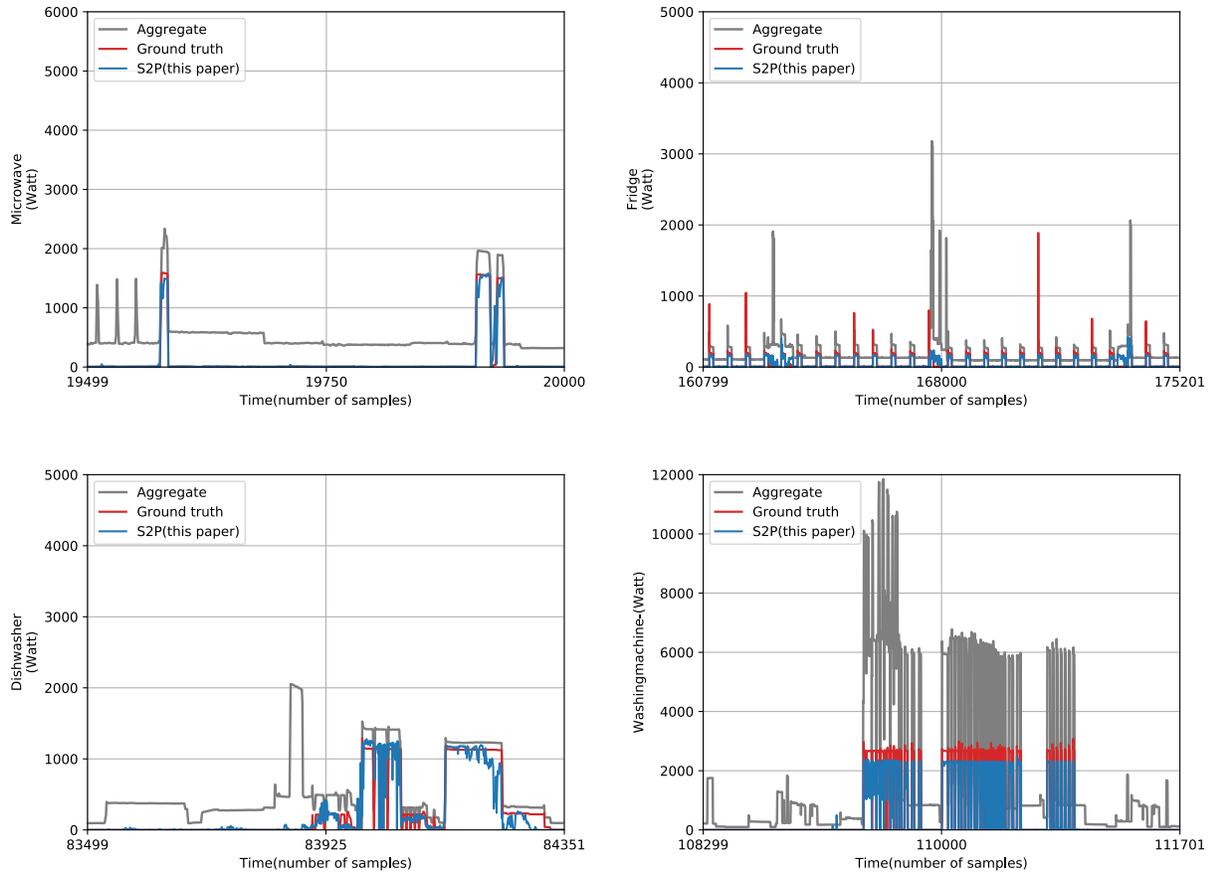

Fig. 6. Detailed load disaggregation of our algorithm on REDD data set.

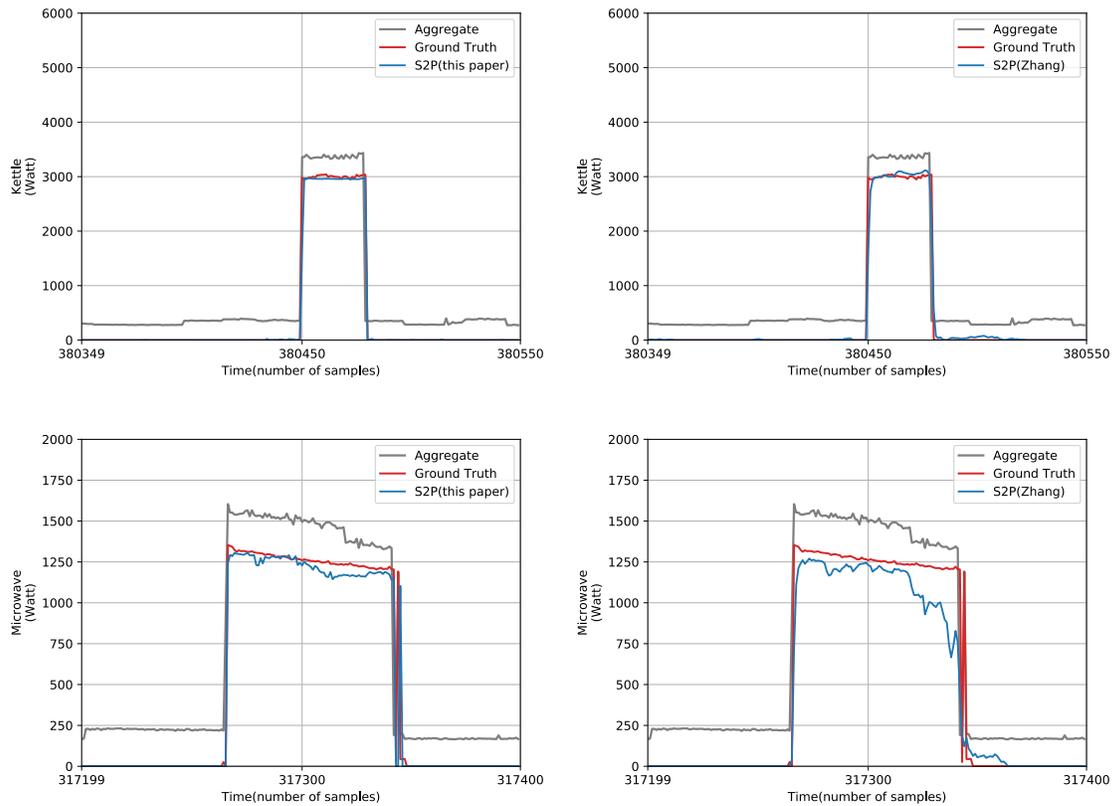

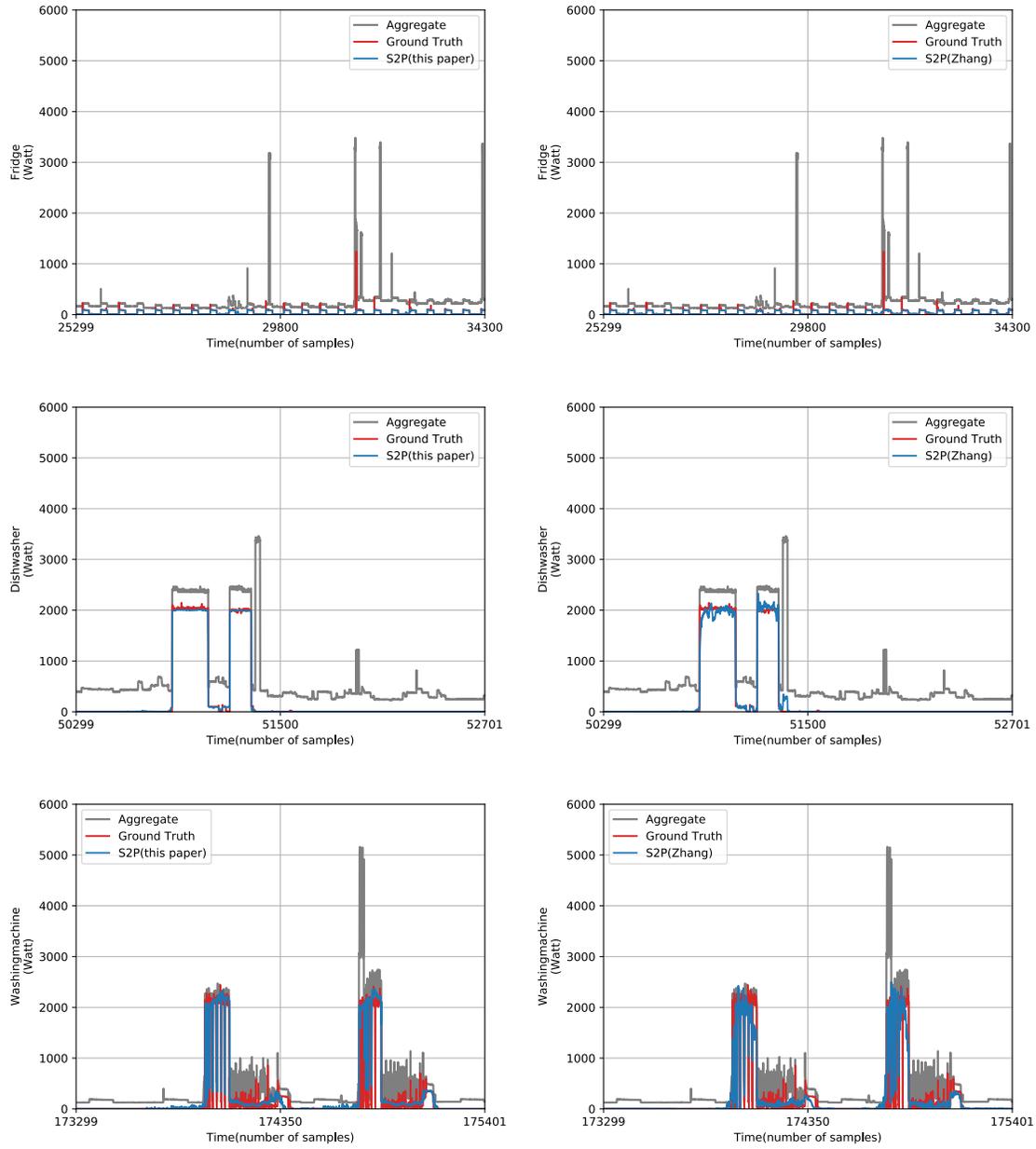

Fig. 7. detailed comparison of load disaggregation between our method and seq2point (Zhang) on household 2 of UK-DALE dataset.

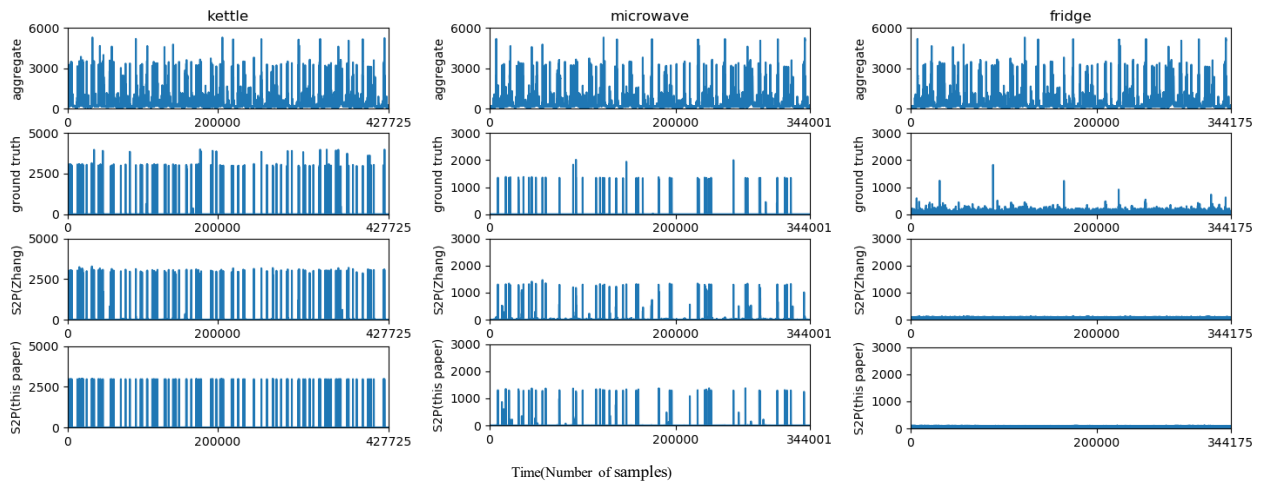

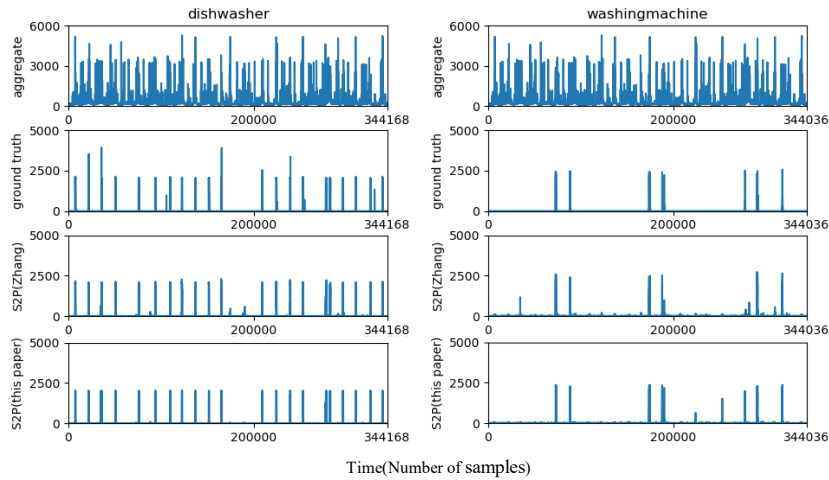

Fig. 8. The comparison of our algorithm with Seq2point (Zhang) on household 2 of UK-DALE dataset.

## 6. Conclusion

All in all, load disaggregation is an important method to achieve smart grid and energy saving. Historical data already proved that traditional disaggregation method is out of date due to its poor accuracy, complicated operation feasibility, and manual features-extract requirements. Deep learning has proved to be a good load decomposition technique. In the paper of Zhang et al., their Seq2point and Seq2seq are superior to the traditional method (AFHMM) and deep learning method like Seq2seq (Kelly) in all appliances. However, in three of the four devices, Seq2point (Zhang) performed better than seq2seq (Zhang). The two experiments demonstrate that our method is superior to seq2point (Zhang) in all devices. So, we make a conclusion that our approach outperforms the existing methods. In addition, our method offers a new way to solve SCBSS problem.

## Acknowledgements

This work was funded by the Natural Science Foundation of China (51767003, 71861002, 61862004) and CERNET Innovation Project (NGII20190310).

The authors would like to thank Multi-Function Computer Center of Guangxi University for its high-performance computer.